\documentstyle{ioplppt}

\newcommand{\ac}{\alpha_{\rm c}}
\newcommand{\vxi}{\vec{\xi}\hspace{0.1em} }
\newcommand{\sgn}{{\rm sgn}} 
\renewcommand{\[}{\begin{equation}} 
\renewcommand{\]}{\end{equation}}
\begin{document}

\jl{1}
\title{An exact learning algorithm for autoassociative neural networks with 
binary couplings}
[Exact learning in binary neuronal networks]

\author{G Milde and S Kobe}

\address{Institut f\"ur Theoretische Physik, Technische Universit\"at 
Dresden, D-01062 Dresden, Germany}

\begin{abstract}

   Exact solutions for the learning problem of autoassociative networks
   with binary couplings are determined by a new method: The use of a
   branch-and-bound algorithm leads to a substantial saving of computing
   time compared to complete enumeration. As a result, fully connected
   networks with up to 40 neurons could be investigated. The network capacity 
   is found to be close to 0.83 .
   
\end{abstract}

\pacs{84.35 75.10Hr}

\nosections The training of neural networks with binary couplings is
believed to belong to the class of NP-complete problems i.e. the average
computing time required to find a solution scales exponentially with the
number of couplings to determine. This exponential scaling is due to the
discrete structure of the space of couplings and is obvious in the case
of complete enumeration. However, theoretical \cite{horner92} and
numerical \cite{patel93} studies showed that it holds as well for
heuristic approaches (e.g. simulated annealing). Training by complete
enumeration has been carried out for small networks with up to 25
neurons \cite{ko89,derrida/griffiths/pruegel-bennet91,uezu/nokura96}.
Heuristic algorithms \cite{patel93,koehler90,fontani/meir91} were used
for networks with up to thousand neurons. Still, the main disadvantage of
heuristic algorithms consists in the uncertainty about the existence of
solutions not found by the algorithm.

Our aim has been to develop an exact algorithm guaranteed to find all
possible solutions in considerably less computing time than complete
enumeration. In \cite{gardner88}, Gardner showed that the space of
interactions in neural network models can be treated in a way similar to
the phase space of spin glass models. Accordingly, it should be possible
to use the branch-and-bound method, already successfully applied to the
search for ground states of a Ising spin glass model
\cite{kobe/hartwig78,klotz/kobe94}, for the training of neural networks
with binary couplings.

Consider an autoassociative network built of $N$ two-state neurons
$s_i=\pm 1 \:(i=1\ldots N)$ and fully connected by binary synaptic
couplings that can take on the values $J_{ij}=\pm 1$. The self couplings
$J_{ii}$ should be set to zero. The task of the network would be to store
a set of patterns $\vxi^\mu\:(\mu=1\ldots p)$ with elements 
$\xi^\mu_i = \pm 1$. A training procedure determines couplings that make
these patterns attractors of the discrete network dynamics
\[ \label{eq:dynamics}
   s_i^{(t+1)} = \sgn \Bigl( \sum_{j=1}^N J_{ij}s_j^{(t)} \Bigr)
   \qquad  i = 1 \ldots N.
\]
The capacity of the network specifies the number of different patterns
that can be stored simultaneously. It is normally expressed as a
critical load $\ac = p_{\rm c}/N$. 

For good retrieval one is interested in large basins of attraction. As
discussed in \cite{ps93a,wong/sherrington90a}, these correspond to large
values of the pattern stability
\[ \label{eq:stability}
    \kappa^\mu = \min_i \Bigl( \frac{1}{\sqrt{N}}
	                       \sum_{j(\ne i)} \xi_i^\mu J_{ij} \xi_j^\mu 
 		        \Bigr).
\]
The maximally stable rule therefore formulates the learning problem as an
optimization task: For a given set of patterns, one has to determine an
optimal set of couplings that maximises the network stability
\[
     \kappa = \min_\mu (\kappa^\mu).
\]
As long as there is no symmetry constraint on the matrix of couplings,
the optimization task separates into the training of $N$ simple
perceptrons with $N-1$ input neurons, corresponding to the individual
rows of the matrix with the self-coupling excluded. The network stability
$\kappa$ emerges as the minimum of the ``perceptron stabilities''
$\kappa_i$.

The new learning algorithm was developed using the branch-and-bound
method, a standard tool of combinatorial optimization theory
\cite{bronstein_E}: To find a row of the matrix of couplings with maximal
stability $\kappa_i$, complete enumeration would check the
$2^{(N-1)}$ possible configurations for optimal ones. Branch-and-bound
starts with a division into a hierarchy of subproblems: each single
coupling is tested with both possible values yet taking into account the
state of the previously (on a trial basis) determined couplings, thus
forming a binary tree of ``incomplete" configurations. Only the final
level of the tree would contain the ``complete" solutions. This division
is the `branching' part of the algorithm. Standing alone it would double
the necessary computing time. Here the `bounding' (and subsequently
cutting) part comes into action: for
each node of the binary tree an upper bound for the best possible
solution of the remaining subproblem is evaluated. Starting point is an
ideal stability, $\kappa_{\rm id} = N-1$, which is obtained if one takes all
terms in the sum (\ref{eq:stability}) to be positive. (Generally, the
maximal stability lies below this ideal stability which can only be
achieved if there is just one pattern to store.) When testing a coupling
$J_{ij}$, this bound will be corrected, taking into account the
already fixed part of the configuration. If it falls under a pre-set
value, e.g. the stability attained by the use of the clipped Hebb rule,
the binary tree is ``cut" at this node, i.e. the subtree of this node
does not need to be considered.  As a result, only a small percentage of
the nodes has to be checked. For $N=25$ we found that only $10^{-4}$ to
$8$ percent of the nodes were evaluated, depending on, e.g., the number of
patterns to store.

Assuming that the evaluation of a node of the binary tree is
approximately as time consuming as checking one possible configuration
during complete enumeration, a comparison of these two methods 
has been done:
As predicted by theory, we still have an exponential scaling of the
algorithm. However, if we set the load $\alpha = p/N$ to 0.5 and look for
one solution with positive stability, the algorithm scales no longer with
$2^N$ but with $2^{0.46 N}$. In the (worst) case of determining all
optimal solutions at $\alpha = 1$, the scaling is $3\times 2^{0.8 N}$.
(That would mean, the algorithm still optimizes a 30-neuron network in
approximately the computing time needed for the complete enumeration of a
25-neuron network.)

We used the branch-and-bound algorithm to determine the capacity of the
network storing random uncorrelated patterns.
Only one row of the coupling matrix was considered
assuming the stability value to be self-averaging in the thermodynamic
limit (cf. \cite{gardner88,ko89}).

The procedure resembles the one used in
\cite{derrida/griffiths/pruegel-bennet91}: For a given value of $N$, the
stability $\kappa^N(\alpha)$ is determined for an increasing number of
patterns until its value becomes negative, signifying that it is no
longer possible to store all patterns. Then the capacity $\ac(N)$ is
determined by a linear interpolation between the last positive
$\kappa^N(\alpha_+)$ and the negative $\kappa^N(\alpha_-)$.  If the
patterns are binary-valued, $\xi_i^\mu = \pm 1$, $\kappa^N(\alpha)$ takes
on discrete values with a spacing of $2/\sqrt{N}$. For $N$ odd, this
discreteness results in two values of $\ac(N)$ corresponding to the first
and last occurrence of $\kappa^N(\alpha) = 0$. This procedure was carried
out for networks with $N=4\ldots 40$. To reduce finite size
(discretization and parity) effects, we also used continuous distributed
patterns (cf. \cite{ko89}). We considered a normalized Gaussian
distribution as well as patterns with elements evenly distributed in the
interval $-1\le\xi_i^\mu \le +1$ (box constraint) to examine the
influence of the pattern distribution.
\begin{figure}
 { \hspace*{-1.2cm}
   \input{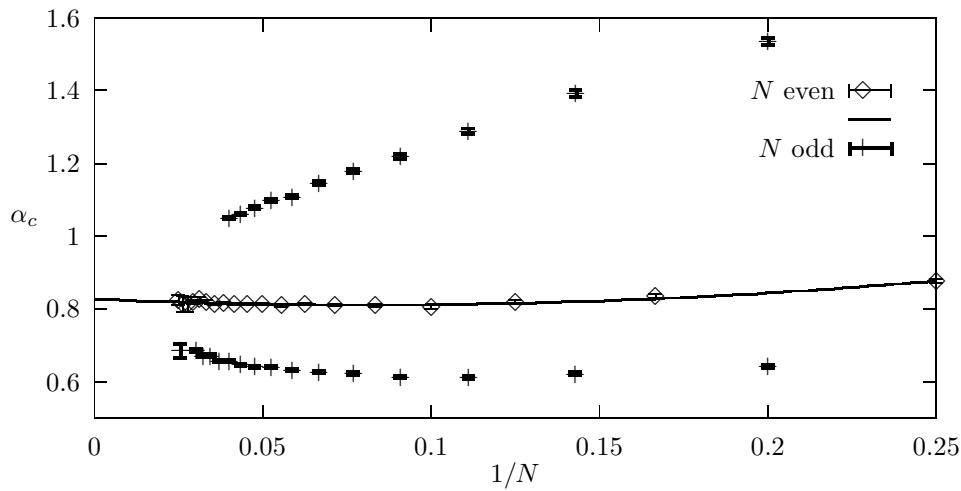} \\[-7mm]
 }
 \caption{\label{fig:ar} Network capacity in the case of random
                         uncorrelated $\pm1$-patterns}
\end{figure}
\begin{figure}
 { \hspace*{-1.2cm}
   \input{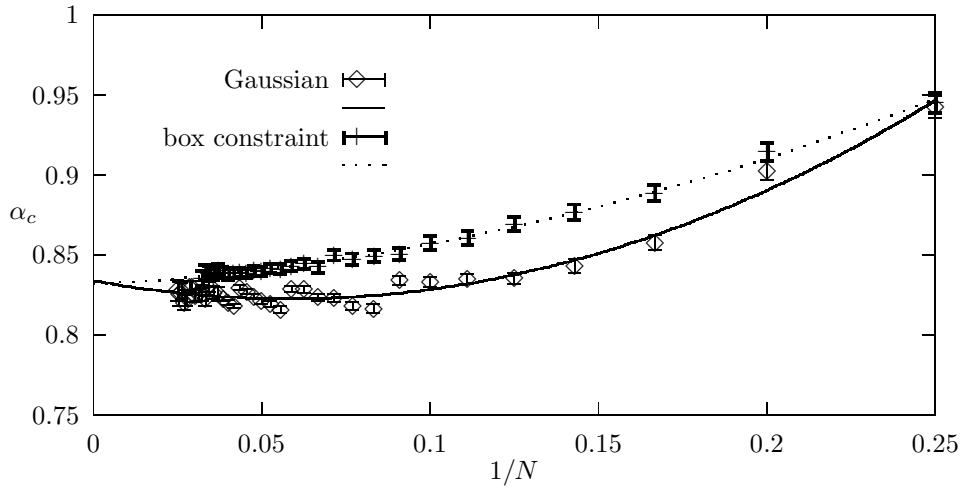} \\[-7mm]
 }
 \caption{\label{fig:arg} Network capacity in the case of continuous 
                          distributed patterns}
\end{figure}
The figures \ref{fig:ar} and \ref{fig:arg} show the dependence of this
capacity on the network size as well as on the nature of the patterns.
The error bars correspond to twice the mean deviation of the average
value (statistical error). Sample size varied between 10 000 for small
systems and 100 for $N=40$. The $\pm1$-patterns exhibit a strong parity
effect which should however vanish in the thermodynamic limit. In
\fref{fig:arg}, the values for the Gaussian patterns show a periodicity
which is probably a result of the linear interpolation as the period of
six corresponds to the passing the zero-line of a stability value
$\kappa^N(\alpha)$. (Remember that the critical capacity is approximately
5/6 and $\alpha$ is restricted to rationals $N/p$). Quadratic fits are
given as a guideline to the eye
(cf. \cite{derrida/griffiths/pruegel-bennet91}).

There is no scaling theory for this problem, however, our numerical
data suggest that the extrapolation to $N\to\infty$ could not be a
linear one. A tentative quadratic extrapolation yields $\ac= 0.834$ for
Gaussian distributed patterns, $\ac = 0.832$ for the box constraint and
$\ac=0.827$ for $\pm1$-patterns and $N$ even. In the case of
$\pm1$-patterns and $N$ odd, a quadratic fit is clearly inadmissible.

A second approach followed the procedure by Krauth and Opper \cite{ko89}
in determining $\kappa_\alpha(N)$ for different values of $\alpha$ and a
subsequent extrapolation to $N \to \infty$. The capacity for
Gaussian distributed patterns is determined as $\ac=0.833$.

We aimed to examine the possibilities and limits of combinatorial
optimization when used for the training of autoassociative neural
networks with binary couplings. The developed branch-and-bound algorithm
allowed us to extend the exact investigation to systems with up to 40
neurons. We were not able to leave the region of strong finite size
effects but could confirm theoretical \cite{krauth/mezard89} and
numerical \cite{ko89,derrida/griffiths/pruegel-bennet91} studies with
additional numerical evidence.
The possibility to determine all solutions of the learning problem also opens
the way for an analysis of the space of solutions similar to the one
already done for the ground states of the Ising spin glass model
\cite{klotz/kobe94}.

\ack Thanks go to W Kinzel for drawing our attention to
the problem, to T Klotz, A Hartwig and J Wei{\ss}barth for many helpful
discussions and to M Opper for additional information. 
GM appreciated a scholarship of the Hans-B\"ockler-Stiftung.

\section*{References}

\end{document}